\documentclass[aps,prl,twocolumn,showpacs, letterpaper]{revtex4}
\usepackage{amsmath, amsthm, amssymb}
\usepackage{graphicx}
\usepackage{bm}

\newcommand{\JS}{\mathrm{JS}}
\newcommand{\kB}{k_\mathrm{B}}
\newcommand{\kT}{\kB T}
\newcommand{\half}{\tfrac{1}{2}}

\newcommand{\x}{ {\mathrm x}}
\newcommand{\J}{\mathop{\mathrm{Jeffreys}}}
\newcommand{\rev}{\tilde}

\begin{document}
\title{The length of time's arrow}
\author{Edward H. Feng}
\affiliation{College of Chemistry, University of California, Berkeley, Berkeley, California, 94720, USA}
\author{Gavin E. Crooks}
\affiliation{Physical Biosciences Division, Lawrence Berkeley National Laboratory, Berkeley, California, 94720, USA}
\date{\today}

\begin{abstract}
An unresolved problem in physics is how the thermodynamic arrow of time 
arises from an underlying  time reversible dynamics.  
We contribute to this issue by developing a measure  of time symmetry breaking, and by using the work fluctuation relations, we determine the time asymmetry of recent single molecule RNA unfolding experiments.  
 We define time asymmetry as the Jensen-Shannon divergence between trajectory probability distributions  of an experiment and its time-reversed conjugate.  
  Among other interesting properties, the length of time's arrow bounds the average dissipation and determines the difficulty of accurately estimating free energy differences in non-equilibrium experiments.
%
%
\end{abstract}

\preprint{}
\maketitle

In our everyday lives  we have the sense that time flows inexorably from the past into the future; water flows downhill;  mountains erode; we are born, grow old and die; we anticipate the future but remember the past. Yet almost all of the fundamental theories  of physics -- classical mechanics, electrodynamics, quantum mechanics, general relativity and so on -- are symmetric with respect to time reversal.  The only fundamental theory that picks out a preferred direction of time is the second law of thermodynamics, which asserts that the entropy of the universe increases as time flows towards the future~\cite{Clausius1865}.  This provides an orientation, or arrow of time, and it is generally believed that all other time asymmetries, such as our sense that future and past are different, are a direct consequence of this thermodynamic arrow~\cite{Eddington1928, Price1996}.

\begin{figure}[t]
\includegraphics[scale=1.0]{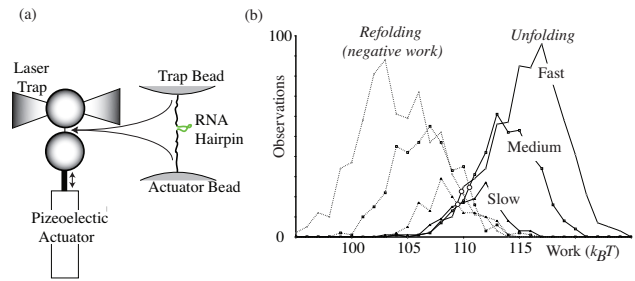}
\caption{
In this Letter, we discuss the definition and measurement of time-asymmetry in microscopic systems. (a) As a concrete example,  we analyze the time asymmetry of a single molecule
 experiment in which an RNA molecule is attached between two beads~\cite{Collin2005}. 
(a)  
 One bead is captured in an optical laser trap that can measure the applied force on the bead. The other bead is fixed to a piezoelectric actuator. The controllable  parameter  $\lambda$ is the distance between the fixed bead and the center of the laser trap. For the forward protocol, the RNA hairpin is initially in thermal equilibrium in the folded state with extension $\lambda(a)$. The extension is then increased to $\lambda(b)$, unfolding the RNA. In the conjugate, time reversed protocol, the RNA is initially in thermal equilibrium in the unfolded state, and the extension is lowered from $\lambda(b)$ back to $\lambda(a)$, allowing the RNA to refold~\cite{Liphardt2002, Collin2005, Bustamante2005,Ritort2006a}.
(b) Histograms of work measurements for folding and unfolding an RNA hairpin at three different rates. Observations are binned into integers centered at 1 $\kB T$ intervals. 
Note that Eq.~(\ref{eq:wft}) predicts that the folding and unfolding work distributions cross at the free energy change. 
 }
\label{fig:exp}
\end{figure}

When the dissipation, or the total increase in entropy, is large, the orientation of time's arrow is self evident. 
If we watch a movie in which shards of pottery jump off the floor, assemble themselves into a cup, and land on a table, then clearly someone has threaded the film through the projector backwards. On the other hand, if the dissipation is microscopic, then the distinction between past and future becomes nebulous.  This is because a more general statement the second law claims the dissipation is positive {\it on average},  $\langle \Delta S_{\text{total}} \rangle \geq 0$~\cite{Evans1993,Jarzynski1997a}.
If we repeat the same experiment many times, the entropy might increase or decrease on different occasions. Only the average dissipation must be positive. Thus, if we view a movie of a microscopic system undergoing a dissipative transformation, we cannot determine with certainty whether time moves forward or backwards. 

Here, we seek a quantitative measure of time asymmetry in a driven microscopic system such as the single molecule RNA pulling experiments explained in Fig.~\ref{fig:exp}. Naively, one might use the average dissipation for this quantification, but we will show that a large average dissipation can arise for dynamics that are essentially time symmetric. 
Instead, we develop a measure based on the Jensen-Shannon divergence between 
the forward and reverse probability distributions of trajectories for a microscopic system. We find that this measure of time's arrow has intuitive physical and information theoretic interpretations and constrains the minimum average dissipation. Moreover, recent advances in far-from-equilibrium statistical physics allow one to measure time's arrow in real world experiments.


\begin{figure*}[t]
\includegraphics[scale=1.0]{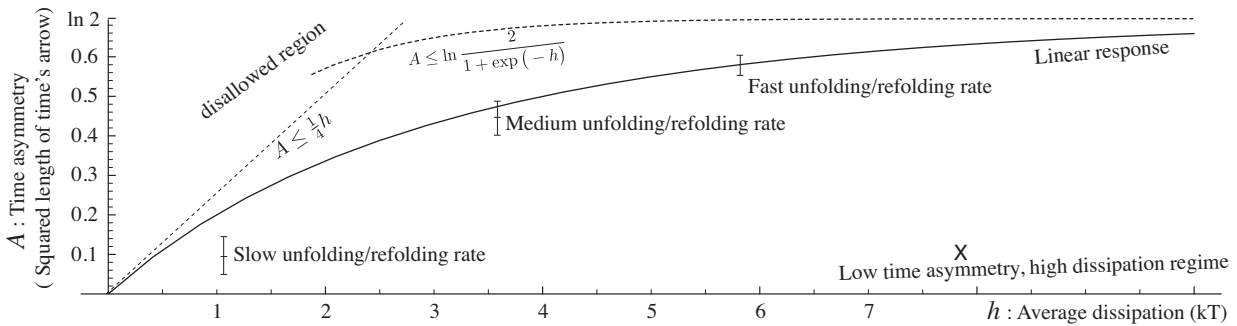} 
\caption{
The squared length of times arrow $A$ [Eq.~(\ref{eq:A})] versus the hysteresis $h$ [Eq.~(\ref{eq:h})], the dissipation averaged across a conjugate pair of forward and reversed experiments.  The experiment is explained in Fig.~\ref{fig:exp}. We equalize the number of data points between conjugate experiments, estimate the free energy from the data, obtain error bars by applying a Bayesian bootstrap~\cite{Rubin1981}, and apply a correction  for experimental errors, as described in~\cite{Maragakis2008b}. The slower the experiment is performed, the closer to thermodynamic reversibility, the lower the dissipation, and the lower the time-asymmetry. 
The  slowest experiments are known to contain the greatest experimental error~\cite{Maragakis2008b}, which may explain the deviation of the slowest data from the linear response trend.
}
\label{fig:Ah}
\end{figure*}

We consider a physical system driven from thermal equilibrium by an external perturbation.
For such an experimental protocol, $\Lambda $ denotes a 
set of
controllable parameters $\lambda(t)  $ for $t \in [a,b] $ which 
describe how the system is driven from the initial equilibrium at
$\lambda(a) $.  
We are also interested in the conjugate time reversed protocol $\rev{\Lambda} $ 
in which
the system begins in thermal equilibrium at $\lambda(b)$ and the controllable parameters retrace the same series of changes, in reverse, back to $\lambda(a)$.
In the single molecule experiments of Fig.~\ref{fig:exp}, the distance between the center of the laser trap and the fixed bead plays the role of $\lambda(t)$.
For each realization of the forward protocol
 $\Lambda $, the system travels along a trajectory~$\x $ which
represents the states $\x(t) $ for $t\in [a,b] $.  We define a conjugate time
reversed trajectory~$\rev{\x} $ such that $\rev{\x} (t) = \x(t) $
for $t\in [b,a] $.



We quantify the intrinsic time asymmetry of a driven system as the distinguishability of conjugate forward and reverse experiments.  
Given a microscopic trajectory $\x $, can we tell if it was generated by the protocol $\Lambda$, or whether it is the time reversal of a trajectory generated by the reverse protocol $\rev{\Lambda} $?  
Specifically, we define the time asymmetry $A$ as 
\begin{equation}
A[\Lambda] \equiv \JS\left( P[\x|\Lambda]\ ;\  P[\rev{\x}|\rev{\Lambda} ] \right)
\label{eq:A}
\end{equation}
in which $P[\x|\Lambda] $ and $P [ \rev{\x}|\rev{\Lambda}] $ are the probabilities of trajectories during 
the forward and reverse protocol respectively, and $\JS$ is the Jensen-Shannon divergence between two probability 
distributions~\cite{Lin1991, Topsoe2000,Endres2003, Osterreicher2003,  Majtey2005}
\begin{equation}
\JS(p;q) = \half  \sum_i p_i \ln \frac{p_i}{\half (p_i+q_i)} +  \half \sum_i q_i \ln \frac{q_i}{\half (p_i+q_i)}\ .
\end{equation}
Each of the two summands is the relative entropy (or
Kullback-Leibler divergence) between one of the distributions 
and the mean of the two distributions.  Hence,  $\JS (p ; q) \geq 0 $, and is equal to zero
only if the two distributions are identical: $p _{i} = q _{i}  $ for all
$i$~\cite{Cover1991}.  The  Jensen-Shannon divergence 
reaches its maximum value of $\ln 2 $ nats [i.e. 1 bit], if the two distributions do not overlap,  
$p _{i} q _{i}   = 0 $  for all $i$, and therefore are perfectly distinguishable.

The Jensen-Shannon divergence has a direct interpretation in terms of a Bayesian inference problem~\cite{Endres2003}.
Suppose we are given a sample $k$ taken from one of two probability distributions, $p$ or $q$.  
With no other way to distinguish between the distributions, the prior probability for the distributions is $P(s )=\{\half , \half \}$ in which $s$ represents either $p$ or $q$. The prior distribution of outcome $k$ is therefore $P(k)  = \half  p_k +   \half  q_k$ while   
the posterior distribution is
\begin{equation}
P(s |k) = \left\{ \frac{\half  p_k }{ \half  p_k  +\half  q_k } , \frac{\half  q_k  }{ \half  p_k  +\half  q_k } \right\} \ .
\end{equation} 
The information gained about $s$ from observing state $k$ is the relative entropy between the posterior and prior distribution. This information averaged over the prior distribution of outcomes is  
\begin{equation}
\langle \Delta I\rangle =   \sum_k P(k)   \Delta I  =  \sum_k P(k)  \sum_{s}  P(s |k) \ln \frac{ P(s  |k) }{ P(s )}\ .
\end{equation}
Some elementary algebra reveals that this average information gain is equal to the Jensen-Shannon divergence, $
\langle \Delta I\rangle 
= \JS(p;q)
$.  Hence, the time asymmetry $A [\Lambda]$ is the average gain in information about the orientation of time's arrow obtained from one realization of the experiment. 
Moreover, the square root of the Jensen-Shannon divergence is a metric between probability distributions~\cite{Endres2003,Osterreicher2003}.  
Consequently, the square root of the time asymmetry $\sqrt{A[\Lambda]}$ measures the distance between the forward and reverse protocols in trajectory space, literally the length of time's arrow.


In addition to its information theoretic interpretation, the time asymmetry can be measured in experiments due to recent advances in far-from-equilibrium statistical physics. 
In particular,  the 
ratio of the probability of a trajectory during 
the forward protocol $P[\x|\Lambda] $ and the probability of its
conjugate trajectory on the reverse protocol $P [ \rev{\x}|\rev{\Lambda}] $ is~\cite{Crooks1998} 
\begin{align}
\label{eq:rev}
\frac{ P[\x|\Lambda] }{ P[\rev{\x}|\rev{\Lambda}] }
& = e^{\beta W [\x| \Lambda] - \beta \Delta F [\Lambda] }
\ ,
\end{align}
in which $\beta=1/\kT$, $T$ is the temperature of the environment in natural units ($\kB$ is the Boltzmann constant), and $W[\x|\Lambda]$ is the work performed on the system during the forward protocol $\Lambda$~\cite{Jarzynski1997a,Crooks1998,Peliti2008}, and  $\Delta F[\Lambda] = F _{ \lambda (b) }- F _{ \lambda (a) } $ is the difference in Helmholtz free energy between the initial and final ensembles.
Eq.~(\ref{eq:rev}) is a direct consequence of the time reversal symmetry of the underlying dynamics\cite{Crooks1998} and implies the work fluctuation theorem 
\begin{equation}
\frac{P ( +W | \Lambda ) }{P  (-W | \Lambda )} = 
e ^{ \beta W - \beta \Delta F } 
\label{eq:wft}
\end{equation}
in which $\Delta F \equiv \Delta F[\Lambda] $.  
%
%
%
%
Moreover, Eq.~(\ref{eq:rev}) gives that the time asymmetry is 
\begin{align}
A[\Lambda]  = 
& \half \left\langle \ln \frac{2}{1 + \exp( - \beta W[\x|\Lambda]+ \beta\Delta F )}  \right\rangle_{\Lambda}
\notag 
\\&+ \half \left\langle \ln \frac{2}{1 + \exp(  -\beta W[\rev{\x}|\rev{\Lambda}] -\beta\Delta F)}  \right\rangle_{\rev{\Lambda}}
\ ,
\label{eq:awork}
\end{align}
a non-linear average of the  forward and reverse dissipation.


Time asymmetry is also closely related to the efficiency with which the free energy can be estimated from non-equilibrium  measurements of the work. To determine $\Delta F $ from experimental realizations of the 
forward and reverse
protocols, Bennett's method gives the log likelihood of the free energy difference as~\cite{Bennett1976,Shirts2003,Maragakis2006, Maragakis2008b}
\begin{align} 
\ell (\Delta F ) &=
\sum_{ i=1 } ^{ K  } 
\ln \frac{1 }{ 1 + e ^{ - \beta W [\x _{ i }|\Lambda] + \beta \Delta 
F } }  
\notag
\\+ &
\sum_{ j=1 } ^{ K  } 
\ln \frac{1 }{ 1 + e ^{ - \beta W [\rev{\x}_{ j }| \rev{\Lambda}] - \beta \Delta 
F  } } \ .
\label{eq:bennett}
\end{align}
Here, $W [\x _{i} |\Lambda ] $ and 
$W [ \rev{\x}_{j}| \rev{\Lambda}]$ are the work measured during the $i$ and  
 $j$ realizations of the forward and  reverse protocols respectively. 
The maximum likelihood estimate $\Delta \hat{F} = \text{argmax}\ \ell(\Delta F)$ has the minimum variance among all
estimators of $\Delta F$~\cite{Shirts2003,Maragakis2006, Maragakis2008b}.  
Comparing Eqs.~(\ref{eq:awork}) and~(\ref{eq:bennett}), the time asymmetry can be estimated with the Bennett likelihood in the large sample limit,~\cite{Crooks2007c,Feng2008}
\begin{equation}
A[\Lambda] \approx  \frac{1}{2K} {\ell}(\Delta F) +\ln2 \ .
\end{equation}
Thus, we can simultaneously estimate  $\Delta F  $ and the time asymmetry by maximizing $A[\Lambda]$ with respect to $\Delta F$.


It is enlightening to contrast the length of time's arrow with
 the hysteresis, the average dissipation of the forward and reverse protocols, 
\begin{equation}
h[\Lambda] =   \half \beta \langle W[\x|\Lambda] \rangle_{\Lambda} +  \half \beta\langle W[\rev{\x}|\rev{\Lambda}] \rangle_{\rev{\Lambda}}\  .
\label{eq:h}
\end{equation}
Because of Eq.~(\ref{eq:rev}), 
the hysteresis is also a divergence between the 
forward and reverse trajectory distributions~\cite{Gaspard2004,Jarzynski2006a, Kawai2007a, Gomez-Marin2008,Andrieux2008}
\begin{equation}
h[\Lambda] =  \half  \J\left(P[\x| \Lambda]\ ;\ P[\rev{\x} |\rev{\Lambda}] \right)
\end{equation}
in which 
\begin{equation}
\J(p;q) = \sum_i p_i \log \frac{p_i}{q_i}  + \sum_i q_i \log \frac{q_i}{p_i}  
\end{equation}
is the Jeffreys J-divergence (or symmetrized Kullback-Leibler  divergence)~\cite{Jeffreys1948, Kullback1951}.
In Fig.~\ref{fig:Ah}, we plot time asymmetry against the
hysteresis for single molecule RNA pulling experiments at  three different rates (see Fig.~\ref{fig:exp}). Both the hysteresis and length of time's arrow increases as the pulling rate increases; however, we will show that this need not always be the case. 
For comparison, we also display the time asymmetry $A[\Lambda] $ in the linear response regime.  The work distributions are normal with variance twice the average dissipation~\cite{Jarzynski1997a}.  

The relative values of  time asymmetry and hysteresis are  bounded by several inequalities. Taneja demonstrated that $A\leq h/4$ using convexity arguments~\cite{Taneja2005}.   
Figure~(\ref{fig:Ah}) shows that this bound
is obeyed by the linear response calculation and the experimental data.
For large values of the hysteresis, we can derive a tighter bound than Taneja. Since the function $f(x)= \ln (1 + e^{-x} )$ is convex, Jensen's inequality~\cite{Cover1991} implies that $\langle \ln (1 + e^{-x} )\rangle \geq  \ln (1 + e^{- \langle x \rangle} )$. Thus,
\begin{align}
&\JS(p;q) 
 = \half  \sum_i p_i \ln \frac{2}{1+ e^{ \ln \frac{q_i}{p_i} }}+ 
\half \sum_i q_i \ln \frac{2}{1+ e^{ \ln \frac{p_i}{q_i}}}
\nonumber
\\
& \leq \half  \ln  \frac{2}{1+ \exp \big(-D(p\|q)\big) }
\notag
+
\half \ln \frac{2}{1+ \exp \big(-D(q\|p)\big) }
\nonumber
\\
& \leq  \ln \frac{2}{1+ \exp\big(- \half \J(p;q)\big) }
\label{eq:in1}
\end{align}            
in which last line follows by a second application of Jensen's inequality. 
Hence, 
$ A \leq  \ln [ 2/ (1+e ^{ -h } ) ] $
which we show in Fig.~\ref{fig:Ah} for large $h$.



However, there is no lower bound to the time asymmetry given the hysteresis.  A system can be almost time symmetric, but exhibit a large average dissipation.  To illustrate this situation, imagine that occasionally, while gently unfolding an RNA hairpin, the RNA becomes stuck in a tangled configuration that resists being pulled apart by force. 
While most repetitions of the experiment give a work measurement very close to the free energy change $W \approx \Delta F $, for the rare instances of entanglement the work is very large $W \gg \Delta F $. 
The normalized distribution of dissipation  $\mathcal{D} = W - \Delta F $ for the forward process may well be approximated as 
\begin{equation}
P (\mathcal{D} | \Lambda ) = (1-p) N (\bar{\mathcal{D}} _{0} ,\sigma _{0} ^{ 2 } ) +
p N (\bar{\mathcal{D} } _{1} ,\sigma _{1} ^{ 2 } )  
\end{equation}
in which $N (\mu,\sigma ^{ 2 } ) $ denotes a normal distribution with mean
$\mu $ and variance $\sigma ^{ 2 }  $. 
The second term corresponds to the rare events, so $p\ll 1 $ and $\bar{\mathcal{D}} _{1} \gg 1 $.
The work fluctuation theorem, Eq.~(\ref{eq:wft}) implies that the dissipation of the reverse process is  
\begin{equation}
P  (\mathcal{D}|\tilde{\Lambda} ) = (1-q) N ( \sigma_0^2 - \bar{\mathcal{D}} _{0} , \sigma_0^2 ) + q N (\sigma_1^2 - \bar{\mathcal{D}} _{1}  ,\sigma _{1} ^{ 2 } )
\end{equation}
in which $q = p \exp({ - \bar{\mathcal{D}} _{1} + \sigma _{1} ^{ 2 } /2 }) $ and normalization requires that
\begin{equation}
(1-p) \exp({ - \bar{\mathcal{D}} _{0} + \sigma _{0} ^{ 2 } /2 }) + p \exp({ - \bar{\mathcal{D}} _{1} + \sigma _{1} ^{ 2 } /2 }) =1
\ . 
\end{equation} 
The linear response regime corresponds to $p=0 $ with $ \bar{\mathcal{D}} _{0} = \sigma _{0} ^{ 2 } /2.$  
For small $p$ and relatively small variance, we almost never see rare trajectories on the reverse protocol with a negative dissipation $- \bar{\mathcal{D} } _{1}  $ since $q$ will be exponentially smaller than $p\ll 1$.  
The values $p=0.1 $, $\bar{\mathcal{D}} _{1} = 150 $, $\bar{\mathcal{D}} _{0} =0.2$ and $\sigma_0^2=1$ give $h=7.8 $ and $A=0.1$ (Marked in Fig~\ref{fig:Ah}), so a system with small time asymmetry can have a large hysteresis.  
While Eq.~(\ref{eq:rev}) shows that dissipation measures the time symmetry breaking of individual trajectories~\cite{Maes2003b, Gaspard2004, Gomez-Marin2008}, the average dissipation is sensitive to unusual events and is not a reliable measure of time asymmetry for the entire system.

One interpretation of relative entropy, and therefore Jeffreys divergence, is that it represents an encoding cost~\cite{Cover1991}. If we encode messages using an optimal code for the message probability distribution $q_i$, but the messages actually arrive with probabilities $p_i$, then each message, on average, will require an  additional $D(p\|q)$ bits to encode compared to the optimal encoding. Analogously, the hysteresis represents a cost, the entropy lost to dissipation.  Thus, the time asymmetry $A$ measures the extent of time-symmetry breaking and the average dissipation measures the price paid. 

The interrelation between time asymmetry and dissipation may be important for molecular motors and other macromolecular biological machinery. One of the central imperatives of any life form is to make tomorrow look  different from today. On the molecular level, this requires rectifying the ever present thermal fluctuations. Since the horizontal-axis in Fig.~\ref{fig:Ah} represents dissipation in units of $\kT$, it takes about 4-8 $\kT$ of free energy per cycle to ensure that a machine mostly advances forward in time, assuming it stays in the linear regime. This is a substantial fraction of the energy budget available from the hydrolysis of an ATP molecule, about 20 $\kT$.

\begin{acknowledgments}
This research was supported by the U.S. Dept. of Energy, under contracts
DE-AC02-05CH11231.  E.H.F.\ thanks the Miller Institute for Basic Research in 
Science for financial support.  We thank Felix Ritort for providing the experimental data used in our analysis.
\end{acknowledgments}

\bibliography{GECLibrary}

\end{document}